\documentclass[prb,aps,twocolumn,nopacs,superscriptaddress,nofootinbib,reprint]{revtex4}
\usepackage{graphicx}
\usepackage{dcolumn}
\usepackage{bm}
\usepackage{amsmath}
\usepackage{epsfig}
\usepackage{bbm}
\usepackage{mathrsfs}
\usepackage{amsfonts}
\usepackage{mathtools}
\usepackage{color} 
\usepackage[colorlinks=true,linkcolor=blue,urlcolor=cyan,citecolor=blue]{hyperref}
\usepackage{multirow} 

\def\ii{{\rm i}}  \def\ee{{\rm e}}

\def\rb{{\bf r}}  \def\Rb{{\bf R}}

\def\kb{{\bf k}}    \def\kparb{{\bf k}_\parallel}
\def\Qb{{\bf Q}}    
\def\me{m_{\rm e}}  

      \def\kp{k_{\rm p}}  
  
\def\dd{{d}}  \def\Kb{{\bf K}}  \def\kbp{{{\bf k}_{\parallel}}}
   \def\Gbp{{{\bf G}_{\parallel}}}
\def\ab{{\boldsymbol{a}}}
    
\def\Gp{{G_{\parallel}}}
  \def\kp{{k_{\parallel}}}
\def\vf{{v_{\text{F}}}}        \def\Phigr{{\Phi^{\rm gr}}}
  \def\kfgr{{k_\text{F}^{\rm gr}}} \def\kfAu{{k_\text{F}^{\rm Au}}}  \def\kfAl{{k_\text{F}^{\rm Al}}}
\def\kfCu{{k_\text{F}^{\rm Cu}}}  \def\kfAg{{k_\text{F}^{\rm Ag}}}
\def\Ef{{E_\text{F}}}  \def\Efgr{{E_\text{F}^{\rm gr}}} \def\EfAu{{E_\text{F}^{\rm Au}}}  \def\EfAl{{E_\text{F}^{\rm Al}}}
        
\def\Vb{{V_{\text{b}}}}        
\def\Efu{{E_{\text{F},1}}}  \def\Efd{{E_{\text{F},2}}}

\begin{document}
\title{Plasmon generation through electron tunneling in twisted double-layer graphene and metal-insulator-graphene systems}

\author{Sandra~de~Vega}
\affiliation{ICFO-Institut de Ciencies Fotoniques, The Barcelona Institute of Science and Technology, 08860 Castelldefels (Barcelona), Spain}
\author{F.~Javier~Garc\'{\i}a~de~Abajo}
\email{javier.garciadeabajo@nanophotonics.es}
\affiliation{ICFO-Institut de Ciencies Fotoniques,  The Barcelona Institute of Science and Technology, 08860 Castelldefels (Barcelona), Spain}
\affiliation{ICREA-Instituci\'o Catalana de Recerca i Estudis Avan\c{c}ats, Passeig Llu\'{\i}s Companys, 23, 08010 Barcelona, Spain}

\date{\today}

\begin{abstract}
The generation of highly-confined plasmons through far-field optical illumination appears to be impractical for technological applications due to their large energy-momentum mismatch with external light. Electrical generation of plasmons offers a possible solution to this problem, although its performance depends on a careful choice of material and geometrical parameters. Here we theoretically investigate graphene-based structures and show in particular the very different performance between (i) two layers of graphene separated by a dielectric and (ii) metal$|$insulator$|$graphene sandwiches as generators of propagating plasmons assisted by inelastic electron tunneling. For double-layer graphene, we study the dependence on the relative tilt angle between the two sheets and show that the plasmon generation efficiency for $4^\circ$ twist angle drops to $\sim20$\% of its maximum for perfect stacking. For metal$|$insulator$|$graphene sandwiches, the inelastic tunneling efficiency drops by several orders of magnitude relative to double-layer graphene, regardless of doping level, metal$|$graphene separation, choice of metal, and direction of tunneling (metal to or from graphene), a result that we attribute to the small fraction of the surface-projected metal Brillouin zone covered by the graphene Dirac cone. Our results reveal a reasonable tolerance to graphene lattice misalignment and a poor performance of structures involving metals, thus supporting the use of double-layer graphene as an optimum choice for electrical plasmon generation in tunneling devices.
\end{abstract}

\maketitle


\section{Introduction}

Surface plasmons --the collective excitations of conduction electrons at the surface of conducting materials-- have been the focus of intense research over the past decades because of a their ability to interact strongly with light, producing huge concentration of electromagnetic energy down to deep-subwavelength regions, accompanied by strong enhancement of the optical field intensity.\cite{SBC10,HLC11} Plasmons are relatively tolerant to defects in the fabrication of the structures supporting them, which are commonly relying on colloid chemistry\cite{L06} and lithography\cite{NLO09} methods, with application to optical sensing,\cite{ZBH14} photocatalysis,\cite{C14} nonlinear optics,\cite{DN07} and optical signal processing,\cite{ZSC06} among other feats. However, the strong spatial confinement produced by optical plasmons has a negative side effect: the coupling cross section from far-field radiation to plasmons is small, typically comparable to or even smaller than the lateral size of the supporting plasmonic nanostructures. Plasmon excitation with light is thus inefficient, which represents a serious obstacle in the development of practical applications. Various methods have been devised to solve this problem, among them the use of funneling structures that focus light down to the spatial extension of the plasmon through the use of gratings\cite{RNE07,AND18} and tips.\cite{FRA12,paper196,BBN11} Quantum emitters have also been employed as a source of plasmons,\cite{NGG11_2} and although they can function as a source of single plasmons, their efficiency is also low.

An alternative to the optical generation of plasmons is provided by electron beams, which have been used in pioneering studies of surface plasmons, including their discovery.\cite{R1957,PS1959,paper149} However, the production and control of electron beams is difficult to combine with integrated devices. Electrons can also inelastically undergo tunneling between neighboring conductors, losing energy that results in the emission of light.\cite{PB92,PPJ15} Recently, plasmon emission produced by inelastic tunneling has been explored in different systems,\cite{WVB10,ZAD13} and this mechanism has also been theoretically explored in sandwiches formed between graphene layers,\cite{SDO16,SDR16,WMC17,paper295} where it is found to be particularly efficient.\cite{paper295}

In this article, we extend a previous study\cite{paper295} and present a comprehensive study of the plasmon-emission efficiency associated with inelastic tunneling between two graphene layers and compare the results with the emission in metal$|$insulator$|$graphene (MIG) configurations. For double-layer graphene (DLG), we assess the dependence of the emission efficiency on the relative graphene twisting angle  and find a reduction of the efficiency when the K points in the two layers are misaligned by more than a few degrees. For MIG configurations, we find plasmon generation rates way below that of DLG.

\section{Spectrally resolved inelastic tunneling current}
\label{theory}

The large plasmon confinement and small thickness compared with the light wavelength in the structures under consideration allow us to work within the quasistatic limit. We thus calculate the probability for an electron to inelastically tunnel between initial and final states $\psi_i$ and $\psi_f$, while transferring an amount of energy $\hbar\omega$ to the materials (e.g., a plasmon), by using the frequency-resolved screened interaction $W(\rb,\rb',\omega)$, which describes the potential created at $\rb$ by a point charge placed at $\rb'$ and oscillating with frequency $\omega$. More precisely, the probability can be written as\cite{paper149,paper295}
\begin{widetext}
	\begin{align}\label{eq:GammaEELS}
		\Gamma(\omega)= 4\dfrac{2e^2}{\hbar} \sum_{i,f} \int \dd^3 \rb \int \dd^3\rb'\, \psi_i^\dagger(\rb)\cdot\psi_f(\rb)  \psi_f^\dagger(\rb') \cdot \psi_i(\rb')\; \text{Im}\{ - W(\rb,\rb',\omega) \}  
		\delta(\varepsilon_f - \varepsilon_i + \omega) \; f_i(\hbar \varepsilon_i) \; [1 - f_f(\hbar \varepsilon_f)] ,
	\end{align}
\end{widetext}
where the leading factor of 4 accounts for spin and valley degeneracies in graphene. This factor is the same for DLG and MIG structures, as they involve at least one graphene layer. Electrons undergo transitions from initial occupied states to final empty states of energies $\hbar \varepsilon_i$ and $\hbar \varepsilon_f$, respectively. The occupation of these levels follows the Fermi-Dirac distributions $f_{i|f}(\hbar\varepsilon_{i|f})=1/\{1 + \exp[(\hbar\varepsilon_{i|f}-E_{{\rm F},i|f})/k_{\rm B}T]\}$, where $E_{{\rm F},i}$ and $E_{{\rm F},f}$ are the corresponding Fermi energies in the emitting and receiving materials, referenced to a common origin of energies. We work here in the $T=0$ limit, for which these distributions become step functions $f_i(\hbar\varepsilon_i)=\Theta(E_{{\rm F},i}-\hbar\varepsilon_i)$ and $1-f_f(\hbar\varepsilon_f)=\Theta(\hbar\varepsilon_f-E_{{\rm F},f})$. For initial or final states in graphene, we recast the sum over electron states as $\sum_{i|f}\rightarrow(2\pi)^{-2}A\int\dd^2\Qb_{i|f}$, while for metals it becomes $\sum_{i|f}\rightarrow(2\pi)^{-3}V\int\dd^3 \kb_{i|f}$, where $A$ and $V$ are the corresponding normalization area and volume, respectively. In the evaluation of Eq.\ (\ref{eq:GammaEELS}), we separate the electron wave functions into parallel and perpendicular components as $\psi(\rb)=\varphi^\parallel(\Rb)\varphi^\perp(z)$, where $\Rb=(x,y)$ are in-plane coordinates. Finally, we present some results below for the loss probability normalized per unit of film surface area $J(\omega)=\Gamma(\omega)/A$ (i.e., the spectrally resolved inelastic tunneling current).

\begin{figure*}
\begin{center}
\includegraphics[width=0.9\textwidth]{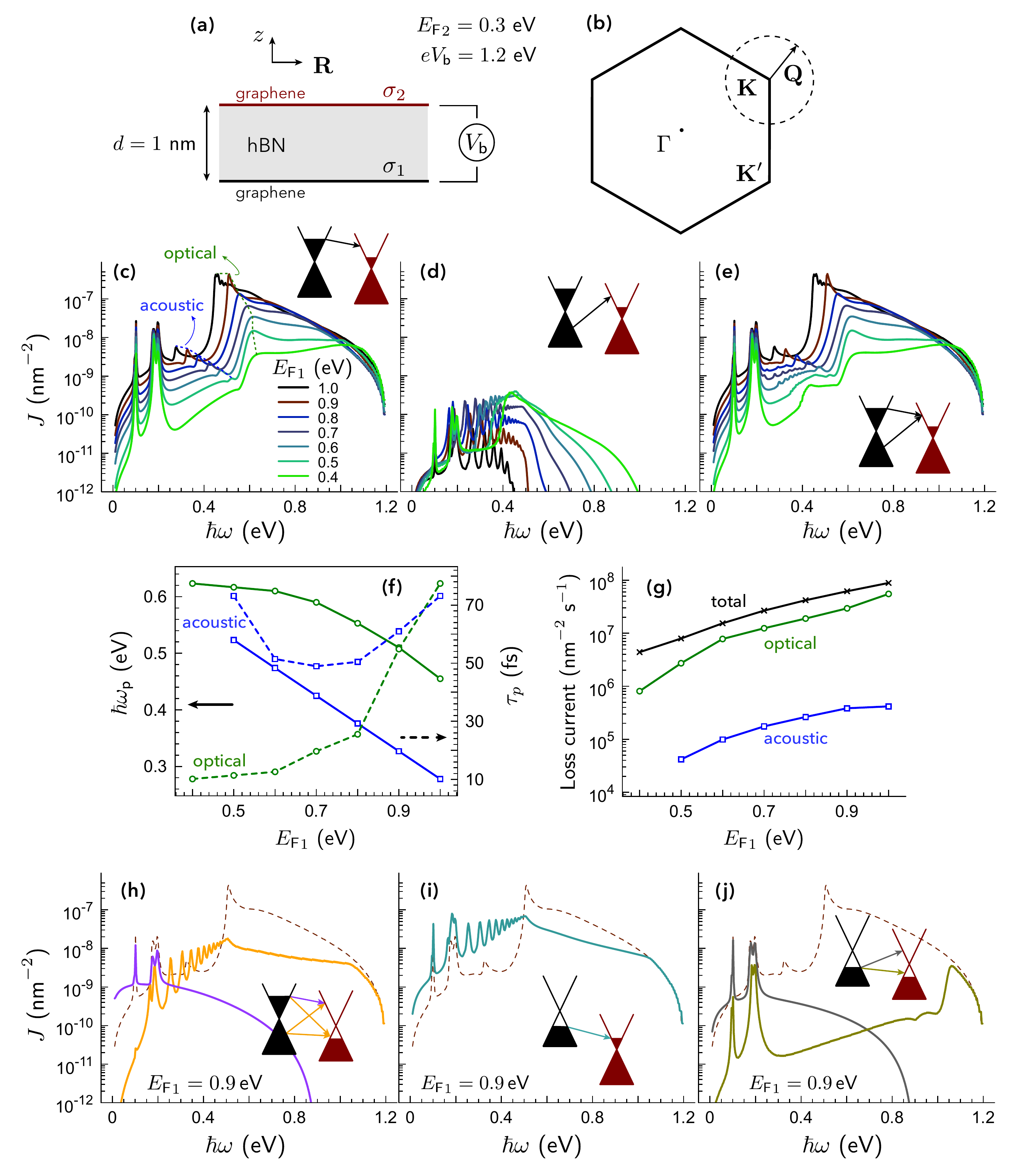}
\caption{{\bf Plasmon generation in double-layer graphene: dependence on doping level.} {\bf (a)} Asymmetrically doped double-layer graphene (DLG, upper layer at Fermi energy $\Efd =0.3$\,eV, fixed in this figure, and lower layer at varying $\Efu $) with an intercalated hBN film ($d=1$\,nm thickness, corresponding to $\sim3$ atomic layer) and gated with a bias voltage $\Vb=1.2$\,eV. {\bf (b)} High symmetry points in the graphene reciprocal lattice. {\bf (c-e)} Spectrally resolved inelastic tunneling current for different values of $\Efu $ when both graphene sheets have electron doping (see insets). We show the separate contributions of (c) conduction-to-conduction and (d) valence-to-conduction transitions, as well as (e) the sum of these two. Optical and acoustic plasmons are marked with labels in (c), while lower-energy, sharp features are associated with the excitation of hBN optical phonons. {\bf (f)} Fermi energy dependence for the energies (left axis, solid curves) and lifetimes (right axis, dashed curves) of the optical (green curves) and acoustic (blue curves) plasmon modes in the DLG structure. {\bf (g)} Dependence of the tunneling electron-current density on $\Efu $, along with the partial contributions of the optical (green curve) and acoustic (blue curve) plasmons. {\bf (h-j)} Spectrally resolved inelastic tunneling current for $\Efu=0.9$\,eV with different types of doping: (h) electron-hole, (i) hole-electron, and (j) hole-hole. Possible transitions between the valence and conduction bands of the two graphene sheets are indicated by arrows in the insets. The contribution of conduction-to-conduction tunneling under electron-electron doping is shown for reference [dashed curves, taken from (c)].}
\label{Fig1}
\end{center}
\end{figure*}

\section{Graphene$|$hBN$|$graphene structures: Dependence on doping level}
\label{GBNGEF}

We first discuss structures formed by two graphene layers separated by a hexagonal boron nitride (hBN) film, assuming perfect alignment between the two graphene reciprocal crystal lattices. We introduced this system in a previous publication,\cite{paper295} but here we study the dependence on the doping level of the two graphene layers. The tunneling structure is sketched in Fig.\ \ref{Fig1}(a): a sandwich of two perfectly stacked graphene layers, with different doping levels, conductivities $\sigma_1$ (bottom) and $\sigma_2$ (top), and connected with a bias voltage $\Vb$, are separated by a distance $d\simeq 1$\,nm ($\sim3$ atomic layers of hBN), which is maintained throughout this work. The first Brillouin zone of the graphene reciprocal lattice is schematically plotted in Fig.\ \ref{Fig1}(b). Around the K point [$\Kb=(2\pi/3a)\,(1,1/\sqrt{3})$, where $a=1.42$\,\AA\, is the nearest-neighbor C-C distance], the electronic band structure is conical in $\Qb$ (the 2D wave vector relative to K) and we can express the electron wave functions as spinors with two components, each of them associated with one of the two carbon atoms in the unit cell.\cite{CGP09} This representation leads to a closed-form expression for the graphene surface conductivity in the random-phase approximation\cite{WSS06,HD07} (RPA), which we use throughout this work with a phenomenological inelastic electron lifetime of 66\,fs (10\,meV energy width, or equivalently, a moderate mean-free path of 66\,nm for a Fermi velocity $v_{\rm F}\approx10^6\,$m/s).

For the evaluation of Eq.\ (\ref{eq:GammaEELS}), we incorporate the following elements:
\begin{itemize}
\item {\it Screened interaction.} We use an analytical expression for the screened interaction describing the graphene$|$hBN$|$graphene sandwich, given in Appendix\ \ref{screenedinteraction} as a generalization of the results presented in Ref.\ \onlinecite{paper295} to an asymmetric environment. This expression incorporates the anisotropy of hBN through its parametrized dielectric function\cite{GPR1966} for in- and out-of-plane directions, which accounts for optical phonons in this material.
\item {\it Parallel electron wave functions.} We use a Dirac-fermion description of the in-plane graphene electron wave functions as
\begin{align}\label{eq:wfi}
\varphi^\parallel_{i|f}(\Rb) = \dfrac{1}{\sqrt{2A}} \;\ee^{\ii \Qb_{i|f}\cdot\Rb} \binom{\ee^{\ii \phi_{i|f} /2}}{\ee^{-\ii \phi_{i|f} /2}} \ee^{\ii \Kb\cdot\Rb},
\end{align}
where $\phi_{i|f}$ is the azimuthal angle of $\Qb_{i|f}$ and $\Kb$ is the wave vector at the K point relative to the $\Gamma$ point [see Fig.\ \ref{Fig1}(b)]. We disregard inter-valley scattering, so each of the two inequivalent K points in the first Brillouin zone produces an identical contribution.
\item {\it Perpendicular electron wave functions.} We describe the evolution of the electron along $z$ through a 1D wave function trapped in the graphene layers by potential wells, which are vertically off-set due to $\Vb$. Full details of this wave function are given elsewhere for graphene,\cite{paper295} while explicit expressions for metals are offered in Appendix\ \ref{electronwave} by treating the surface in the one-electron step-potential approximation.
\end{itemize}

The hybridization of plasmons in the two neighboring layers produces a characteristic hybridization scheme, leading to two plasmon branches: optical and acoustic. Acoustic plasmons have lower energy and their out-of-plane electric-field profile looks antisymmetric. In contrast, optical plasmons possess higher energy and symmetric field profiles, so they are generally easier to excite and manipulate, and therefore, more suitable for photonic applications.

Tunneling requires $\Vb\neq 0$ and $\Efu\neq\Efd$. We take $\Efu >\Efd  $, and hence, favor tunneling from layer 1 (bottom) to layer 2 (top). We explore the effect of varying the doping-level difference between the two graphene layers by calculating the spectrally resolved inelastic tunneling current shown in Fig.\ \ref{Fig1}(c), where we fix $\Efd =0.3$\,eV and $e\Vb=1.2$\,eV. Energy splitting due to phonons in the hBN film can be observed at low energy transfers (below 0.3\,eV), losing strength as the Fermi energy difference $\Delta\Ef=\Efu-\Efd$ decreases.

Acoustic plasmons become apparent for $\Delta\Ef>0.2\,$eV. These resonances have an average plasmon lifetime of $\sim$50-70\,fs [Fig.\ \ref{Fig1}(f), blue-dashed curve] and undergo a redshift with increasing $\Delta\Ef$ [Fig.\ \ref{Fig1}(f), blue-solid curve]. In contrast, optical plasmons become more pronounced when $\Efu > 2\Efd$, they experience a milder redshift with increasing $\Delta\Ef$ [Fig.\ \ref{Fig1}(f) green-solid curve], and their lifetime is rapidly increasing from $\sim$10\,fs to $\sim$80\,fs [Fig.\ \ref{Fig1}(f) green-dashed curve].

We quantify the fraction of the total inelastic current invested into exciting plasmons by comparing the area under the whole spectrum in Fig.\ \ref{Fig1}(g) for different values of $\Efu$ [Fig.\ \ref{Fig1}(g), black curve] to the area under either the optical (green curve) or acoustic (blue curve) plasmon regions. The average generation efficiency for the selected values of $\Vb$ and $\Efd$ lies in the $\sim20$-60\% range for optical plasmons and $< 1$\% for acoustic modes.

\begin{figure*}
\begin{center}
\includegraphics[width=1\textwidth]{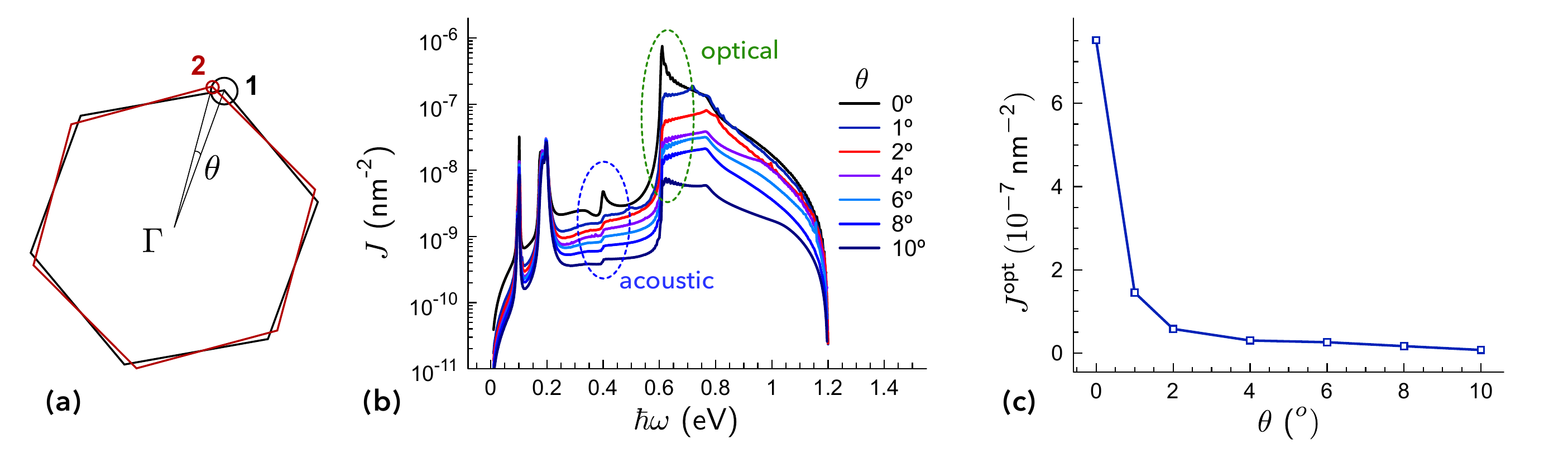}
\caption{{\bf Plasmon generation in double-layer graphene: dependence on tilt angle.} {\bf (a)} First Brillouin zones of two graphene layers tilted by an angle $\theta$ ($5^\circ$ in this plot), along with their respective Fermi surfaces (circular cross sections of their respective Dirac cones, we only plot one per layer) for Fermi energies of $\Efu =1$\,eV and $\Efd =0.5$\,eV. {\bf (b)} Spectrally resolved inelastic tunneling current for different tilt angles and the following choice of parameters: $e\Vb$= 1.2\,eV, $\Efu $=1\,eV, $\Efd $=0.5\,eV, and $d$=1\,nm [see Fig.\ \ref{Fig1}(a)].  Optical and acoustic plasmons are marked with labels, while lower-energy, sharp features are associated with the excitation of hBN optical phonons. {\bf (c)} Decay of the partial contribution of optical-plasmon-excitation to the inelastic current at $\hbar\omega\approx0.61\,$eV as a function of tilt angle.} 
\label{Fig2}
\end{center}
\end{figure*}

\section{Graphene$|$hBN$|$graphene structures: Dependence on twist angle}

In practical devices, the alignment of the two graphene layers can be a challenge, so we examine the effect of a finite twist angle between their respective lattices. Rotations in real space result in rotations around the $\Gamma$ point in momentum space. In Fig.\ \ref{Fig2}(a) we depict two reciprocal lattices corresponding to the bottom (1, black) and top (2, red) layers for a finite rotation angle $\theta$. The circles surrounding their respective K points represent the projections of their Fermi surfaces on 2D momentum space for Fermi energies $\Efu =1$\,eV and $\Efd =0.5$\,eV. Inelastic tunneling requires the two circles to be either overlapping or closer to each other than the plasmon momentum, which is small compared with the Fermi momenta ($k_{\rm F}=\Ef/\hbar v_{\rm F}$, i.e., the radii of the plotted circles), leading to a cutoff angle $\sim10^\circ$.
 
We introduce the twist angle $\theta$ in our formalism through the electron wave function of the final states, so that we maintain the initial states as in Eq.\ (\ref{eq:wfi}) with $\Kb=\Kb_i$, but write the final wave function as
\begin{align}\label{eq:wff}
\varphi^\parallel_f(\Rb) = \dfrac{1}{\sqrt{2A}} \;\ee^{\ii \Qb_f\cdot\Rb} \binom{\ee^{\ii (\phi_f - \theta) /2}}{\ee^{-\ii (\phi_f - \theta) /2}} \ee^{\ii \Kb_f\cdot\Rb}.
\end{align}
Notice that $\Kb_i$ and $\Kb_f$ are the K points corresponding to the centers of the circles in Fig.\ \ref{Fig2}(a). The spinor product in Eq.\ (\ref{eq:GammaEELS}) involving the parallel wave functions thus becomes
\begin{align}
&{\varphi^\parallel_i}^\dagger(\Rb)\cdot\varphi^\parallel_f(\Rb)  {\varphi^\parallel_f}^\dagger(\Rb') \cdot \varphi^\parallel_i(\Rb') \nonumber\\
&= \dfrac{1}{2A^2}\, \ee^{\ii (\Kb_f-\Kb_i+\Qb_f-\Qb_i)\cdot(\Rb - \Rb')} \,[1 + \cos(\phi_i - \phi_f + \theta)],  \nonumber
\end{align}
with $ \Kb_f - \Kb_i = (2\pi/3a)\,(\cos\theta - \sin\theta/\sqrt{3} - 1,\,\sin\theta+\cos\theta/\sqrt{3}-1/\sqrt{3})$. Introducing this expression together with the parallel-wave-vector decomposition of $W$ [Eq.\ (\ref{Wkpar})] into Eq.\ (\ref{eq:GammaEELS}) and performing the $\Rb$ and $\Rb'$ integrals, we find the condition $\kbp+\Kb_f-\Kb_i+\Qb_f-\Qb_i=0$, which guarantees momentum conservation. We find it convenient to separate the contribution of the perpendicular wave functions as
\begin{align}
	I_1(\kp,\omega) = - \int \dd z\int \dd z'\; &{\varphi^\perp_i}^\dagger(z)\varphi^\perp_f(z)  {\varphi^\perp_f}^\dagger(z')\varphi^\perp_i(z') \nonumber\\
&\times\text{Im}\{  W(\kp,z,z',\omega)\}, \nonumber
\end{align}
where $\varphi^\perp_{i|f}$ are solutions corresponding to the two quantum wells that we use to model the graphene layers.\cite{paper295} Putting these elements together, and considering the initial electron to tunnel from the conduction band of layer 1, we can write from Eq.\ (\ref{eq:GammaEELS}) the spectrally-resolved tunneling current density
\begin{align}
J(\omega)&=\frac{\Gamma(\omega)}{A}= \dfrac{e^2}{4\pi^4} \int\dd^2\Qb_i \int\dd^2\kbp \; I_1(\kp,\omega) \label{eq:Jw_rot_gBNg}\\
&\times\delta(\hbar\vf Q_f -\hbar\vf Q_i +\Efu-\Efd- e\Vb+ \hbar\omega) \nonumber \\
&\times [1+\cos(\phi_i-\phi'_f-\theta)] \nonumber\\
&\times\Theta(\Efu -\hbar \vf Q_i)\;\Theta(\hbar\vf Q_f -\Efd), \nonumber
\end{align}
where $\Gbp=\Kb_f-\Kb_i+\kbp$, $\phi'_f=\tan^{-1}[(Q_{yi}-\Gp_y)/(Q_{xi}-\Gp_x)]$, and $Q_f=|\Qb_i-\Gbp|$. We note that the Dirac $\delta$-function ensures energy conservation in Eq.\ (\ref{eq:Jw_rot_gBNg}), limiting the spectral range to $-\vf\Gp+e\Vb/\hbar<\omega<\vf\Gp+e\Vb/\hbar$. For $\theta=0$ (no twist), this expression reduces to that of Ref.\ \onlinecite{paper295}, where incidentally, a spurious factor of $2\pi$ was accidentally introduced in the numerical results.

We represent in Fig.\ \ref{Fig2}(b) the tunneling current obtained from Eq.\ (\ref{eq:Jw_rot_gBNg}) for different rotation angles when fixing the bias voltage to $\Vb=1.2$\,eV and the Fermi energies to $\Efu =1$\,eV and $\Efd =0.5$\,eV. When plotting the maximum of $J(\omega)$ associated with the optical plasmon ($\hbar\omega_p^\text{opt}\simeq0.61$\,eV), we find a sharp decay with increasing twist angle [Fig.\ \ref{Fig2}(c)], although we still maintain $\sim20$\% of the maximum value for $\theta=4^\circ$, further indicating a reasonable tolerance of this type of device against unintended misalignments below the $1^\circ$ level.

So far, we have considered electron doping in both of the graphene layers. We can straightforwardly repeat the above analysis to find expressions that apply to situations in which one or both of the graphene layers is doped with holes. Furthermore, for sufficiently high $\Vb$, electrons can tunnel from the valence band of layer 1 even when both layers have electron doping, which leads to an additional term in the integrand of Eq.\ (\ref{eq:Jw_rot_gBNg}); although we give the resulting expression below these lines, we do not consider such high voltages in this study. We find
\begin{align}
J(\omega)= \dfrac{e^2}{4\pi^4} &\int\dd^2\Qb_i \int\dd^2\kbp I_1(\kp,\omega) \nonumber\\
              &\times [1+\cos(\phi_i-\phi'_f-\theta)] \;\Delta(\Qb_i,\kparb,\omega), \nonumber
\end{align}
where
\begin{widetext}
\begin{align}
\Delta(\Qb_i,\kparb,\omega) =
\left\{\begin{array}{ll}
\;\;\,\delta(\hbar\vf Q_f -\hbar\vf Q_i +\Efu-\Efd- e\Vb+ \hbar\omega)
                                          \;\Theta(\Efu -\hbar \vf Q_i) \;\Theta(\hbar\vf Q_f -\Efd) & \text{(e-e doping)} \\
+\delta(\hbar\vf Q_f +\hbar\vf Q_i +\Efu-\Efd- e\Vb+ \hbar\omega) \;\Theta(\hbar\vf Q_f -\Efd), & \\ & \\
\;\;\,\delta(-\hbar\vf Q_f -\hbar\vf Q_i + \Efu+\Efd - e\Vb+ \hbar \omega)
                                          \;\Theta(\Efu - \hbar \vf Q_i) \;\Theta(\Efd - \hbar\vf Q_f) & \text{(e-h doping)}\\
+\delta(\hbar\vf Q_f  -\hbar\vf Q_i + \Efu+\Efd - e\Vb+\hbar \omega)
                               \;\Theta(\Efu - \hbar \vf Q_i) & \\
+\delta(-\hbar\vf Q_f +\hbar\vf Q_i + \Efu+\Efd - e\Vb+ \hbar \omega)
                               \;\Theta(\Efd - \hbar\vf Q_f) & \\
+\delta(\hbar\vf Q_f  +\hbar\vf Q_i + \Efu+\Efd - e\Vb+\hbar \omega), & \\ & \\
\;\;\,\delta(\hbar\vf Q_f   +\hbar\vf Q_i - \Efu-\Efd - e\Vb+ \hbar \omega)
                                 \;\Theta(\hbar \vf Q_i - \Efu) \;\Theta(\hbar\vf Q_f - \Efd), & \text{(h-e doping)} \\ & \\
\;\;\,\delta(-\hbar\vf Q_f   +\vf Q_i - \Efu+\Efd - e\Vb+ \hbar \omega) \;\Theta(\hbar \vf Q_i - \Efu) \;\Theta(\Efd - \hbar\vf Q_f) & \text{(h-h doping)} \\
+\delta(\hbar\vf Q_f   +\hbar\vf Q_i - \Efu+\Efd - e\Vb+ \hbar \omega) \;\Theta(\hbar \vf Q_i - \Efu), &
\end{array} \right.
\nonumber
\end{align}
\end{widetext}
the labels e (electron) and h (hole) indicate the type of doping in the first and second layers, respectively, and several $\delta$-functions appear in each expression depending on whether the electron originates in the valence or conduction bands (for electron doping in layer 1) and whether the electron tunnels to the valence or conduction bands (for hole doping in layer 2). We define $\Efu$ and $\Efd$ as positive quantities, although it is understood that they represent a lowering of the Fermi energy relative to the Dirac point when doping with holes.

In Figs.\ \ref{Fig1}(d,h-j) we show results for two perfectly stacked graphene layers with different types of doping. Arrows in the insets indicate the different tunneling channels. Additionally, we compare the results to a reference spectrally-resolved tunnelling probability associated with conduction-to-conduction transitions in two electron-doped graphene layers, which we conclude in fact to be the most effective mechanism to generate plasmons under the conditions of the figure.

\section{Plasmon generation in metal$|$insulator$|$graphene tunneling structures}

\begin{figure*}
\begin{center}
\includegraphics[width=1\textwidth]{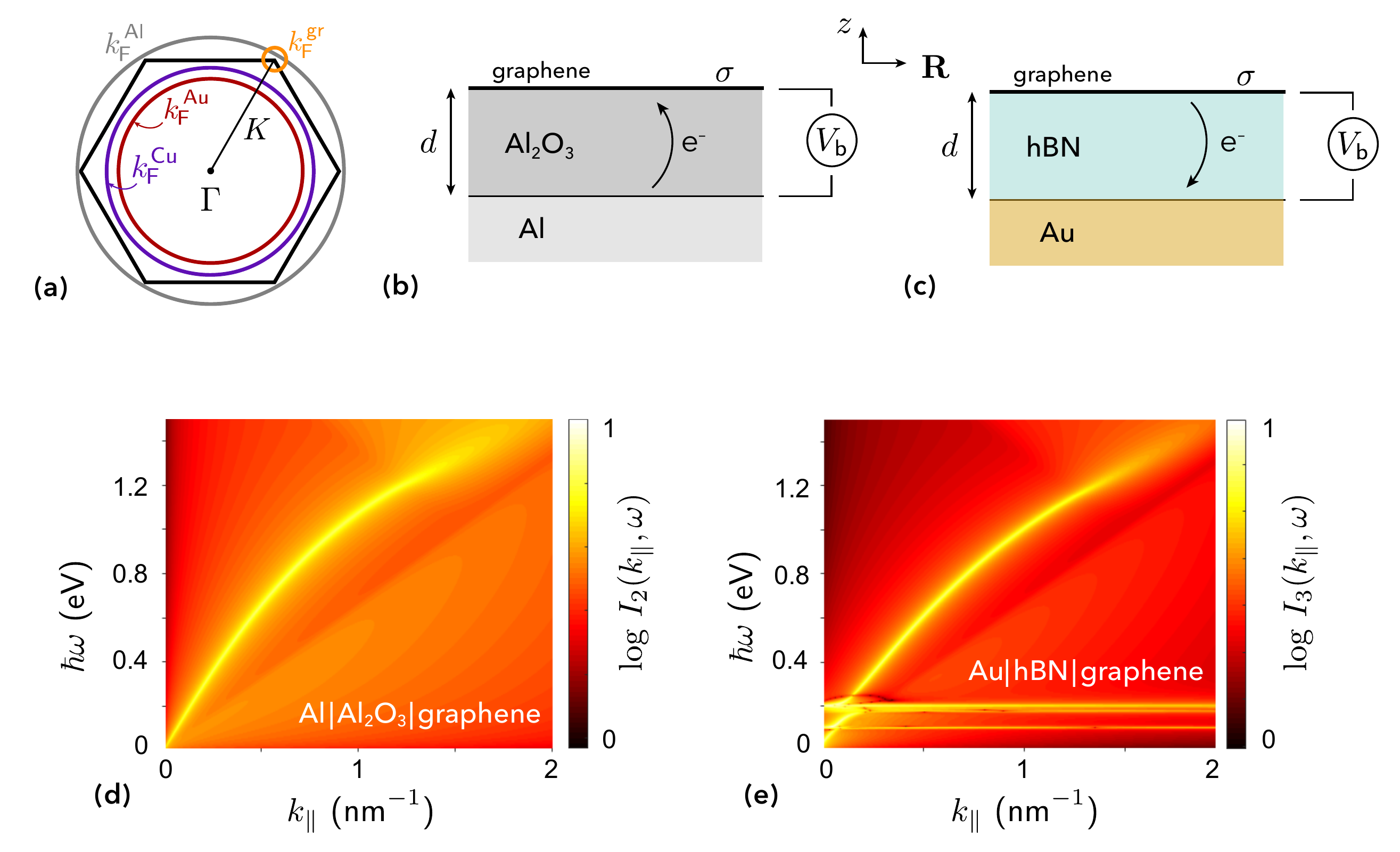}
\caption{{\bf Plasmon generation in metal$|$insulator$|$graphene tunneling structures}. {\bf (a)} Projection of the Fermi surface on the space of wave vectors parallel to the graphene plane for Al (grey), Au (red), and Cu (violet), along with the first Brillouin zone of graphene and its Fermi surface for $\Efgr $=1\,eV doping. {\bf (b)} Scheme of the Al$|$Al$_2$O$_3|$graphene structure gated to a bias voltage $\Vb$. The thickness of the oxide layer is $d=1\,$nm. The tunneling current goes from Al to graphene. {\bf (c)} Scheme of the Au$|$hBN$|$graphene structure gated to a bias voltage $\Vb$. The thickness of the hBN layers is also $d=1\,$nm ($\sim$\,3 layers). The tunneling current goes from graphene to Au. {\bf (d)} Plasmon dispersion relation given by the poles of the screened potential for the Al$|$Al$_2$O$_3|$graphene structure (1\,eV doping, tunneling from aluminum to graphene) when the metal is approximated as a perfect conductor [see Eq.\ (\ref{eq:I1_Ali})]. {\bf (e)} Same as (d) with Au instead of Al and electrons tunneling from graphene to the metal [see Eq.\ (\ref{eq:I1_Alf})].} 
\label{Fig3}
\end{center}
\end{figure*}

We now compare the performance of DLG and MIG structures. In the latter, one of the graphene layers is substituted by a semi-inifinite metal medium. When the tunneling electrons go from the metal to the graphene sheet, not all metals can comply with momentum conservation. In Fig.\ \ref{Fig3}(a) we represent the surface projection of the Fermi spheres (assuming for simplicity an independent-electron description of the metal Fermi sea) of different plasmonic metals (Au, Cu, and Al) compared with the first Brillouin zone of graphene, where we further show the graphene Fermi surface for a doping level $\Efgr=1$\,eV. When comparing the Fermi wave vectors of these metals ($\kfAu =12.1$\,nm$^{-1}$, $\kfAg=12.0$\,nm$^{-1}$, $\kfCu=13.6$\,nm$^{-1}$, $\kfAl =17.5$\,nm$^{-1}$) with the wave vector at the graphene K point ($K=17.0$\,nm$^{-1}$), considering the small radius of the graphene Fermi circle ($\kfgr=\Efgr/\hbar\vf=1.52$\,nm$^{-1}$), we find that only the Al Fermi sea overlaps the graphene Dirac cone, so this is the only one among the good plasmonic metals in which electrons can tunnel to graphene (without the mediation of phonons or defects). Then, it is reasonable to consider the structure represented in Fig.\ \ref{Fig3}(b), consisting of a graphene layer on top of an aluminum surface coated with a 1\,nm layer of oxide (Al$|$Al$_2$O$_3|$graphene). A gate voltage is then introduced to make metal electrons tunnel into the doped graphene sheet and excite plasmons in the MIG structure.

In contrast, when electrons tunnel from graphene to the metal, the situation is reversed, so it is favorable to have the occupied Dirac-cone region outside the projected metal Fermi sea. This situation is encountered with different choices of metal, and in particular with gold, separated from graphene by 3 atomic layers of hBN, as depicted in Fig.\ \ref{Fig3}(c) (Au$|$hBN$|$graphene).

For simplicity, we calculate the screened interaction (see Appendix\ \ref{screenedinteraction}) assimilating the metal to a perfect electric conductor ($|\epsilon|\rightarrow\infty$ is a good approximation for Au and Al within plasmonic energy range under consideration). We further approximate the response of aluminum oxide by using a constant isotropic permittivity $\epsilon_{\text{Al}_2\text{O}_3}=3$ (the measured permittivity\cite{P1985} only changes by $\sim4$\% within the energy range under consideration). Finally, we use the RPA conductivity for graphene\cite{WSS06,HD07} (see Sec.\ \ref{GBNGEF}) and the anisotropic permittivity described in 	Appendix\ \ref{screenedinteraction} for hBN.\cite{GPR1966}

We calculate the spectrally-resolved tunneling probability using Eq.\ (\ref{eq:GammaEELS}) and the formalism described in Sec.\ \ref{theory}, with the metal wave functions (either as initial or final states, depending on the configuration) described as
\begin{align}
\psi_{i|f}(\rb)= \varphi^\perp_{i|f}(z,k_{i|f\,z}) \dfrac{1}{\sqrt{A}}\, \ee^{\ii\kb_{i|f}^\parallel\cdot \Rb}.\nonumber
\end{align}	
The perpendicular component of these wave functions depend on the incident wave vector along $z$ (i.e., $k_{i|f\,z}$) according to the explicit expressions derived in Appendix\ \ref{electronwave} under the assumption of a step potential to represent the metal$|$insulator interface.

\begin{figure*}
\begin{center}
\includegraphics[width=1\textwidth]{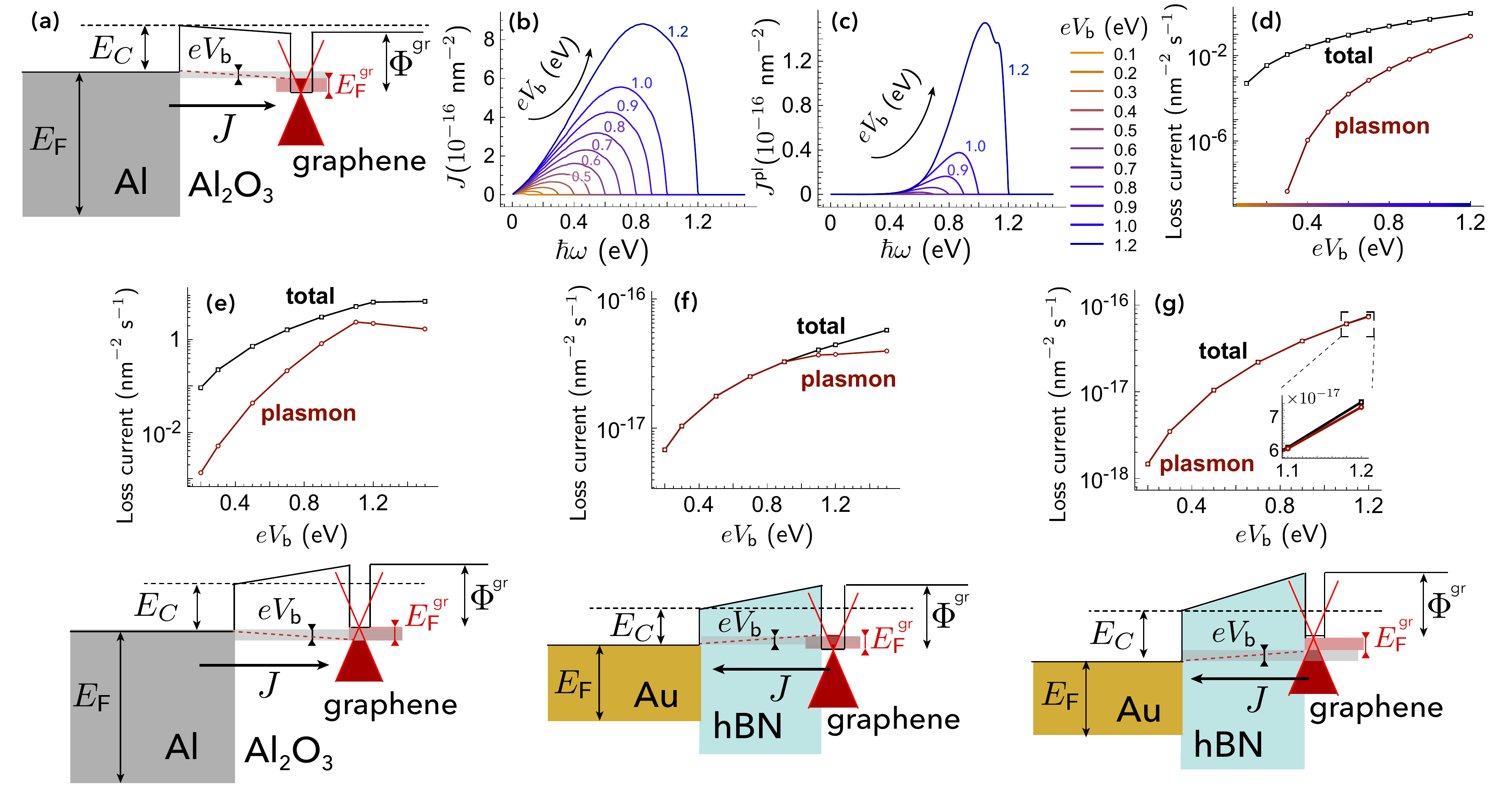}
\caption{{\bf Overview of plasmon generation via electron tunneling in MIG structures}. {\bf (a)} Energy bands to scale for the Al$|$Al$_2$O$_3|$graphene sandwich when graphene is electron-doped to $\Efgr=1$\,eV. These parameters hold for the other configurations shown in this figure. {\bf (b,c)} Spectrally resolved inelastic tunneling current for $\Efgr $=1\,eV and different bias voltages for the system of (a): (b) total current and (c) plasmon-excitation contribution [using Eq.\ (\ref{eq:I1pli})]. The value of $e\Vb$ for each curve coincides with its spectral cutoff energy $\hbar\omega$. {\bf (d)} Dependence of the total inelastic tunneling current (black) and partial plasmon contribution (red) on bias voltage after frequency integration of Eq.\ (\ref{eq:Jw_Al-gre}) for the system of (a). {\bf (e-g)} Inelastic tunneling currents equivalent to those in (d) for the configurations shown in the lower insets: (e) Al-Al$_2$O$_3$-hole-doped graphene, (f) Au-hBN-electron-doped graphene and (g) Au-hBN-hole-doped graphene, as obtained from the frequency integral of Eqs.\ (\ref{eq:Jw_Al-grh}), (\ref{eq:Jw_gre-Au}) and (\ref{eq:Jw_grh-Au}), respectively. The graphene work function is $\Phigr =4.7$\,eV and we choose a bias $e\Vb=0.7$\,eV in the sketches.} 
\label{Fig4}
\end{center}
\end{figure*}

As graphene has two sublattices, a small phase difference $\ee^{\ii\kb_i^\parallel\cdot{\bf a}}$ has to be introduced to account in the coupling to the metal wave functions. Additionally, depending on the applied bias, the graphene doping level, and the type of doping (i.e., electrons or holes), only one or the two graphene Dirac cones can be engaged in the tunneling process. Taking these elements into account, the expression for the inelastic tunneling current calculated from Eq.\ (\ref{eq:GammaEELS}) needs to be specified for each of the following MIG configurations:
\begin{widetext}
\begin{itemize}
\item Al$\rightarrow$Al$_2$O$_3\rightarrow\,$electron-doped-graphene [see Fig.\ \ref{Fig4}(a)]
\begin{align}
J(\omega) = \dfrac{\hbar e^2}{4\pi^5} &\int\dd^2\Qb_f \int\dd^3\kb_i \; I_2(\kp,k_{iz},\omega) \;\Theta\left(\EfAl  - \hbar^2 k_i^2/2\me \right)\left[ 1 + \cos\left(\phi_f + \kb_i^\parallel\cdot\ab \right) \right]  \label{eq:Jw_Al-gre} \\
&\times\delta\left(\hbar\vf Q_f - e\Vb - \Efgr - \hbar^2 k_i^2/2\me + \EfAl   + \hbar\omega \right) \; \Theta(\hbar \vf Q_f - \Efgr), \nonumber
\end{align}
\item Al$\rightarrow$Al$_2$O$_3\rightarrow$hole-doped-graphene [see Fig.\ \ref{Fig4}(e)]
\begin{align}
J(\omega) = \dfrac{\hbar e^2}{4\pi^5} &\int\dd^2\Qb_f \int\dd^3\kb_i \; I_2(\kp,k_{iz},\omega) \;\Theta \left(\EfAl  - \hbar^2 k_i^2/2\me \right)\left[ 1 + \cos\left(\phi_f + \kb_i^\parallel\cdot\ab \right) \right]  \label{eq:Jw_Al-grh} \\
&\times\bigg[ \delta\left(-\hbar\vf Q_f - e\Vb + \Efgr - \hbar^2 k_i^2/2\me + \EfAl   + \hbar\omega \right) \; \Theta (\Efgr-\hbar \vf Q_f) \nonumber \\
&+ \delta\left(\hbar\vf Q_f - e\Vb + \Efgr - \hbar^2 k_i^2/2\me + \EfAl   + \hbar\omega \right)\bigg], \nonumber
\end{align}
\item electron-doped-graphene$\rightarrow$hBN$\rightarrow$Au [see Fig.\ \ref{Fig4}(f)]
\begin{align}
J(\omega) = \dfrac{\hbar e^2}{4\pi^5} &\int\dd^2\Qb_i \int\dd^3\kb_f \; I_3(\kp,k_{fz},\omega) \;\Theta \left( \hbar^2 k_f^2/2\me - \EfAu  \right)\left[ 1 + \cos\left(\phi_i + \kb_f^\parallel\cdot\ab \right) \right]  \nonumber \\
&\times\bigg[ \delta\left(-\hbar\vf Q_i - e\Vb + \Efgr + \hbar^2 k_f^2/2\me - \EfAu   + \hbar\omega \right) \; \Theta (\Efgr-\hbar \vf Q_i) \label{eq:Jw_gre-Au} \\
&+ \delta\left(\hbar\vf Q_i - e\Vb + \Efgr + \hbar^2 k_f^2/2\me - \EfAu   + \hbar\omega \right)\bigg], \nonumber
\end{align}
\item hole-doped-graphene$\rightarrow$hBN$\rightarrow$Au [see Fig.\ \ref{Fig4}(g)]
\begin{align}
J(\omega) = \dfrac{\hbar e^2}{4\pi^5} &\int\dd^2\Qb_i \int\dd^3\kb_f \; I_3(\kp,k_{fz},\omega) \;\Theta\left( \hbar^2 k_f^2/2\me -\EfAu   \right)\left[ 1 + \cos\left(\phi_i + \kb_f^\parallel\cdot\ab \right) \right]  \label{eq:Jw_grh-Au} \\
&\times\delta\left(\hbar\vf Q_i - e\Vb - \Efgr  + \hbar^2 k_f^2/2\me - \EfAu   + \hbar\omega \right) \; \Theta(\hbar \vf Q_i - \Efgr ), \nonumber
\end{align}
\end{itemize}
where the arrows indicate the direction of electron tunneling, and we have defined
\begin{align} \label{eq:I1_Ali}
	I_2(\kp,k_{iz},\omega) = -\int \dd z\int \dd z'\; {\varphi^\perp_i}^\dagger(z,k_{iz})\varphi^\perp_f(z)  {\varphi^\perp_f}^\dagger(z')\varphi^\perp_i(z',k_{iz})\; \text{Im}\left\{W(\kp,z,z',\omega)\right\}
\end{align}
for metal$\rightarrow$graphene tunneling and
\begin{align} \label{eq:I1_Alf}
I_3(\kp,k_{fz},\omega) = -\int \dd z\int \dd z'\; {\varphi^\perp_i}^\dagger(z)\varphi^\perp_f(z,k_{fz})  {\varphi^\perp_f}^\dagger(z',k_{fz})\varphi^\perp_i(z')\; \text{Im}\left\{W(\kp,z,z',\omega)\right\}
\end{align}
for graphene$\rightarrow$metal. We note that energy conservation can only be fulfilled for energies $\hbar\omega<e\Vb$ in all of the these structures.
\end{widetext}

The contour plots in Fig.\ \ref{Fig3}(d,e) portray Eqs.\ (\ref{eq:I1_Ali}) and (\ref{eq:I1_Alf}) for fixed values of $k_{iz}$ and $k_{fz}$ in the Al$|$Al$_2$O$_3|$graphene and Au$|$hBN$|$graphene structures, respectively, with a graphene Fermi energy fixed to $\Efgr =1$\,eV. They clearly reveal a plasmon mode arising from the poles of the screened potential $W$. Additionally, Fig.\ \ref{Fig3}(e) shows low-energy features associated with hBN phonons in the Au$|$hBN$|$graphene structure. 

In order to quantify the amount of inelastic current associated with plasmon generation, we separate the screened interaction $W=W^\text{dir}+W^\text{ref}$ into the sum of external ($W^\text{dir}$) and reflected ($W^\text{ref}$) components (see Appendix\ \ref{screenedinteraction}). Plasmons arise from the poles of $W^\text{ref}$, originating in a denominator of the form $\eta(\kp)=1+4\pi\ii\sigma\kp/\omega+\tilde\epsilon_2+\ee^{2qd} \left(\tilde\epsilon_2-1-4\pi\ii\,\sigma\kp/\omega\right)$, where $\tilde\epsilon_2=\sqrt{\epsilon_{2x}\epsilon_{2z}}$, $q=\kp\sqrt{\epsilon_{2x}/\epsilon_{2z}}$, $\epsilon_{2x}$ and $\epsilon_{2z}$ are the permittivities of the insulator (Al$_2$O$_3$ or hBN) along in-plane ($x$) and out-of-plane ($z$) directions, and $\sigma$ is the graphene conductivity. We now isolate this pole and write $W^\text{ref}=\eta^{-1} \tilde W^\text{ref}$. In the vicinity of the plasmon pole $\kp=k_{\rm p}$, we can Taylor-expand $\eta (\kp)=\eta'(k_p)(\kp-k_{\rm p})$ to first order and approximate the plasmon contribution to Eqs.\ (\ref{eq:I1_Ali}) and (\ref{eq:I1_Alf}) as
\begin{widetext}
\begin{align}
I_2^\text{pl}(\kp,k_{iz},\omega) &= -\int \dd z\int \dd z'\; {\varphi^\perp_i}^\dagger(z,k_{iz})\varphi^\perp_f(z)  {\varphi^\perp_f}^\dagger(z')\varphi^\perp_i(z',k_{iz})\text{Im}\left\{\dfrac{\tilde W(\kp,z,z',\omega)}{\eta'(k_p)(\kp - k_p)}\right\}, \label{eq:I1pli}\\
I_3^\text{pl}(\kp,k_{fz},\omega) &= -\int \dd z\int \dd z'\; {\varphi^\perp_i}^\dagger(z)\varphi^\perp_f(z,k_{fz})  {\varphi^\perp_f}^\dagger(z',k_{fz})\varphi^\perp_i(z')\text{Im}\left\{\dfrac{\tilde W(\kp,z,z',\omega)}{\eta'(k_p)(\kp - k_p)}\right\}. \label{eq:I1plf}
\end{align}
\end{widetext}
By using either Eqs.\ (\ref{eq:I1_Ali})-(\ref{eq:I1_Alf}) or Eqs.\ (\ref{eq:I1pli})-(\ref{eq:I1plf}) into Eqs.\ (\ref{eq:Jw_Al-gre})-(\ref{eq:Jw_grh-Au}), we obtain the total spectrally-resolved inelastic tunneling current $J(\omega)$ or the contribution arising from plasmon generation $J^\text{pl}(\omega)$, respectively.

We show the calculated spectrally-resolved currents $J(\omega)$ and $J^\text{pl}(\omega)$ in Fig.\ \ref{Fig4}(b,c) for Al$\rightarrow$Al$_2$O$_3\rightarrow$electron-doped-graphene inelastic tunneling with a bias voltage in the 0.1\,eV $<e\Vb<$1.2\,eV range and $\Efgr $=1\,eV. A significant spectral broadening in the total current [Fig.\ \ref{Fig4}(b)] originates in the availability of multiple inelastic channels associated for example with electron-hole pair generation in the metal. Indeed, plasmons make a relatively moderate contribution  [see Fig.\ \ref{Fig4}(c)], as we conclude by comparing the total to the plasmon-based $\omega$-integrated inelastic currents [Fig.\ \ref{Fig4}(d)]. The plasmon generation efficiency (number of plasmons per tunneling electron, obtained from the ratio $J^\text{pl}/J$) varies from $<10^{-6}$ for $\Vb<0.3\,$eV to 0.1 for voltages close to 1.2\,eV, where the plasmon generation rate is $\sim10^{-2}$\,nm$^{-2}$s$^{-1}$. Figure\ \ref{Fig4}(e-g) represents the total and plasmon-based inelastic currents for the configurations depicted below the graphs, from which we conslude that Al$\rightarrow$Al$_2$O$_3\rightarrow$hole-doped-graphene exhibits better performance, with plasmon generation rates approaching $\sim1\,$nm$^{-2}$s$^{-1}$, while for Au$|$hBN$|$graphene structures the currents are 16 orders of magnitude smaller. Overall, comparing the performance of these MIG systems to DLG, we find the latter to be much more efficient, with inelastic currents reaching 10$^8$\,nm$^{-2}$s$^{-1}$ (see Fig.\ \ref{Fig1} and Ref.\ \onlinecite{paper295}).

\section{Conclusions}

Our simulations reveal the suitability of DLG heterostructures as plasmon sources even under moderately unfavorable conditions produced by a finite twist angle between the two graphene sheets, with $>70$\% efficiency for a twist angle as large as $4^\circ$. Additionally, we find the relation $\Efu =2\Efd $ between the Fermi energies of the two layers to be an optimum choice to maximize the plasmon emission rate of both acoustic and optical plasmons, although the efficiency is still in the $>10$\% range for order-one variations in the Fermi energies. Furthermore, our study of MIG structures leads to efficiencies that are orders of magnitude lower than DLG. 

From an intuitive viewpoint, these conclusions can be understood using the following argument. We are trying to project electrons from one electrode into the other, which requires conservation of parallel momentum, differing by just the plasmon momentum, which is small compared with the size of the Brillouin zone. For DLG, the matching is most efficient with perfect alignment, involving large overlaps of their respective Dirac cones, while the tolerance against twisting mentioned above can be roughly quantified by the angle required to produce total mismatch between the Dirac cones, of the order of a few degrees for doping levels of 0.5-1\,eV. Unfortunately, when one of the electrodes is a metal, although the conduction electron density at the surface is 2-3 orders of magnitude larger than that of charge carriers in graphene, they are distributed over a larger momentum-space region, rendering the overlap with the Dirac cone smaller; this effect, together with a weaker plasmon strength in MIG compared with DLG, results in much poorer plasmon generation rates.

In conclusion, DLG offers an optimum choice for the electrical generation of plasmons based upon inelastic electron tunneling, which is moderately robust against twisting of the layers (i.e., misalignment of their respective K points in the first Brillouin zone) and takes place for broad ranges of doping (i.e., it is also tolerant with respect to variations in doping).

\appendix

\section{Electron wave functions in metal$|$insulator interfaces}
\label{electronwave}

We approximate the electron wave functions in the metal along the $z$-direction as those of a quantum step-potential, assuming the configuration shown in the following scheme (metal for $z<0$ and insulator for $z>0$):
\begin{center} \includegraphics[width=0.25\textwidth]{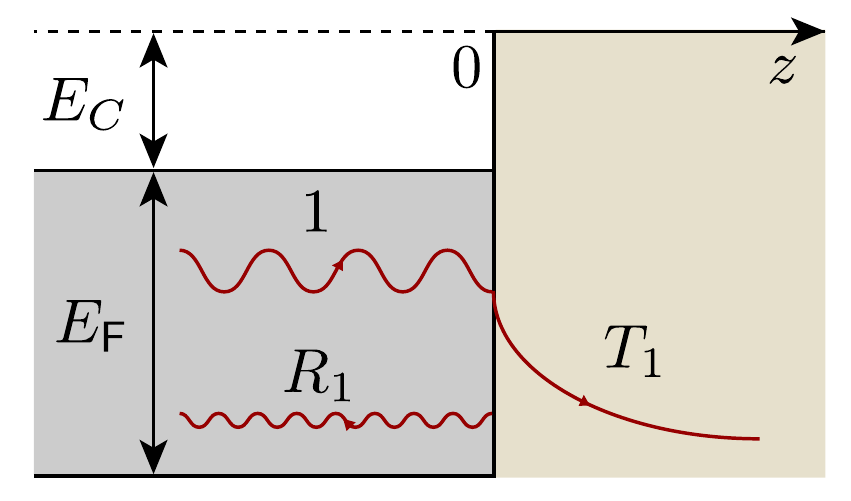} \end{center}
The Fermi energies of the metals here considered are $\EfAl =11.7\,$eV and $\EfAu =5.53\,$eV. The parameter $E_C$ accounts for the band gap of the insulators, for which we take $E_C^{\text{Al}_2\text{O}_3}=3.5\, $eV\cite{FK15} and $E_C^\text{hBN}=2.6\,$eV (i.e., half of their gap energies, under the assumption that their Fermi levels are in the center of the gap, aligned with that of the metal). Moreover, $T_1$ and $R_1$ are the transmission and reflection coefficients at the interface, respectively. By solving the time independent Schr\"odinger equation $(-\hbar^2/2\me) d^2\varphi^\perp(z)/dz^2 + V(z) \varphi^\perp(z) = E \varphi^\perp(z)$ with $V(z)=-V_0 \theta(-z)$ and $V_0=E_C+\Ef$, an electron incident on the interface from the metal side has the wave function
\begin{align}
\varphi^\perp_i(z,k_{iz})=\left\{ \begin{array}{ll}
\dfrac{1}{\sqrt{2\pi}} \left(\ee^{\ii k_{iz} z} + R_1\ee^{-\ii k_{iz} z}\right), & \quad z<0\\
\dfrac{1}{\sqrt{2\pi}} T_1 \ee^{- k_2 z}, & \quad z>0\\
\end{array} \right.
\nonumber
\end{align}
where $k_{iz}$ is the electron wave vector along the interface normal in the metal side, $k_2 = \sqrt{2 m V_0/\hbar^2 - k_{iz}^2}$, $ R_1 = (k_{iz} - \ii k_2)/(k_{iz} + \ii k_2) $ and $T_1=2k_{iz} / (k_{iz} +\ii k_2)$.

\section{Screened interaction potential}
\label{screenedinteraction}

Because of the 2D translational invariance of the system, we can write the screened interaction as
\begin{align}
W(\rb,\rb',\omega) = \int \dfrac{\dd^2\kbp}{(2\pi)^2} \, \ee^{\ii\kbp\cdot(\Rb-\Rb')}\, W(\kp,z,z',\omega)
\label{Wkpar}
\end{align} 
in terms of its parallel wave-vector components. We consider a multilayer structure with the permittivities $\epsilon_{1|2|3}$ and surface conductivities $\sigma_{1|2}$ as defined by the following scheme:
\begin{center} \includegraphics[width=0.25\textwidth]{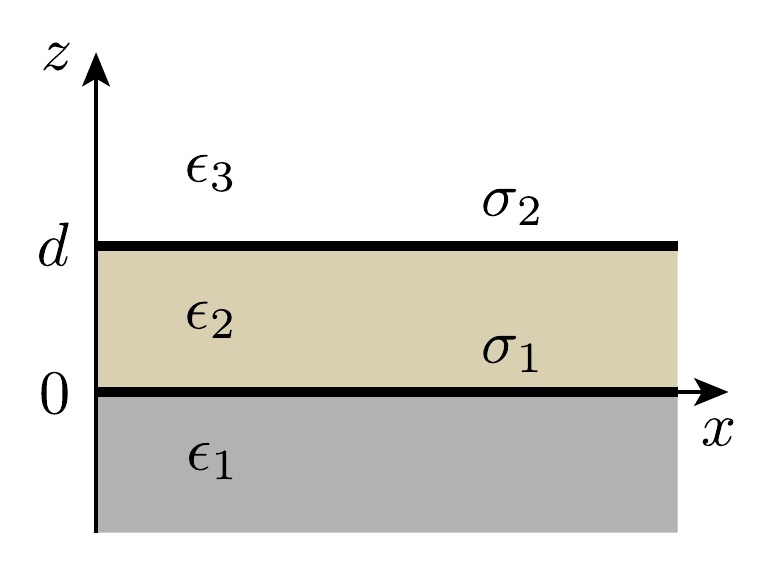} \end{center}
This structure can describe the two configurations considered in this work: MIG (with on of the $\sigma_j$'s set to zero) and DLG. An expression for $W$ was presented in a previous work\cite{paper295} for a symmetric environment ($\epsilon_1=\epsilon_3=1$), which we generalize here for the asymmetric configuration sketched above. Using the methods explained in that work, we separate
\begin{align}
W(\kp,z,z',\omega)=W^{\rm dir}(\kp,z,z',\omega)+W^{\rm ref}(\kp,z,z',\omega),
\nonumber
\end{align}
into direct and surface-reflection contributions, where
\begin{widetext}
\begin{align}
W^{\rm dir}(\kp,z,z',\omega)=\frac{2\pi}{\kp}\times
\left\{\begin{array}{ll}
(1/\epsilon_1)\ee^{-\kp|z-z'|},  &\quad z,z'<0 \\
(1/\epsilon_3)\ee^{-\kp|z-z'|},  &\quad z,z'>0 \\
(1/\tilde{\epsilon}_2)\ee^{-q|z-z'|},  &\quad 0<z,z'<d \\
0, &\quad \text{otherwise}
\end{array} \right.
\nonumber
\end{align}
and
\begin{align}
&W^{\rm ref}(\kp,z,z',\omega)=\frac{(2\pi/\kp)}{1-A'_1A'_2\ee^{-2qd}}
\nonumber\\
&\times
\left\{\begin{array}{lll}
(1/\epsilon_3)\ee^{\kp(2d-z-z')}\left[A_2+A'_1(A_2+B'_2)\,\ee^{-2qd}\right],  &d<z, &d<z' \\ 
(1/\epsilon_3)B_2\;\ee^{\kp(d-z)}\left[\ee^{-q(d-z')}+A'_1\;\ee^{-q(d+z')}\right],  & d<z, &0<z'<d \\ 
(\tilde{\epsilon}_2/\epsilon_1) B_1B_2\;\ee^{\kp (d-z+z')}\ee^{-qd},  & d<z, &z'<0 \\ 
(1/\epsilon_3)B_2\;\ee^{\kp(d-z')}\left[\ee^{-q(d-z)}+A'_1\;\ee^{-q(d+z)}\right],  & 0<z<d, &d<z' \\ 
(1/\tilde{\epsilon}_2)\big\{
A'_1\;\ee^{-q(z+z')}+A'_2\;\ee^{-q(2d-z-z')} & & \nonumber\\
\quad\;\;\;
+A'_1A'_2\left[\ee^{-q(2d+z-z')}+\ee^{-q(2d-z+z')}\right]
\big\},  \quad\quad & 0<z<d, &0<z'<d \\ 
(1/\epsilon_1)B_1\;\ee^{\kp z'}\left[\ee^{-qz}+A'_2\;\ee^{-q(2d-z)}\right],  & 0<z<d, &z'<0 \\ 
(\tilde{\epsilon}_2/\epsilon_3) B_1B_2\;\ee^{\kp(d+z-z')}\ee^{-qd},  &z<0, &d<z'\\ 
(1/\epsilon_1)B_1\;\ee^{\kp z}\left[\ee^{-qz'}
+A'_2\;\ee^{-q(2d-z')}\right],  &z<0, &0<z'<d\\ 
(1/\epsilon_1)\ee^{\kp(z+z')}\left[A_1+A'_2(A_1+B'_1)\,\ee^{-2qd}\right],  &z<0, &z'<0
\end{array} \right.
\nonumber
\end{align}
\end{widetext}
This expression is applicable when the material in medium 2 is anisotropic (e.g., hBN) by defining $\tilde{\epsilon}_2=\sqrt{\epsilon_{2x}\epsilon_{2z}}$ as the geometrical average of the in-plane ($x$) and out-of-plane ($z$) permittivities, with the square root chosen to yield a positive imaginary part. Also, $q=\kp\sqrt{\epsilon_{2x}/\epsilon_{2z}}$ (with ${\rm Re}\{q\}>0$) is the effective out-of-plane wave vector in that medium, and we have defined the coefficients
\begin{align}
B_1&=2\epsilon_1/(\epsilon_1+\tilde{\epsilon}_2+\beta_1), \nonumber\\
B'_1&=(\tilde{\epsilon}_2/\epsilon_1) B_1, \nonumber\\
B_2&=2\epsilon_3/(\epsilon_3+\tilde{\epsilon}_2+\beta_2), \nonumber\\
B'_2&=(\tilde{\epsilon}_2/\epsilon_3) B_2, \nonumber\\
A_j&=B_j-1, \nonumber\\
A'_j&=B'_j-1, \nonumber\\
\beta_j&=4\pi\ii\kp\sigma_j/\omega, \nonumber
\end{align}
for $j=1,2$. In this work, we approximate the permittivity of alumina as a constant value $\tilde\epsilon_2=\epsilon_{\text{Al}_2\text{O}_3}=3$, we use the perfect-conductor limit for metals ($|\epsilon_1|\rightarrow\infty$), we describe the graphene conductivity in the RPA,\cite{WSS06,HD07} and we use the expression \cite{GPR1966}
\begin{align}\nonumber
\epsilon_{2\ell}(\omega)\approx\epsilon_{\infty,\ell}+\sum_{i=1,2}s_{\ell i}^2/[\omega_{\ell i}^2-\omega(\omega+\ii\gamma_{\ell i})],
\end{align}
for the conductivity of hBN, where
$\epsilon_{\infty,z}=4.10$,
$s_{z 1}=70.8\,$meV,
$\omega_{z 1}=97.1\,$meV,
$\gamma_{z 1}=0.99\,$meV,
$s_{z 2}=126\,$meV,
$\omega_{z 2}=187\,$meV,
and
$\gamma_{z 2}=9.92\,$meV
for the $z$ component ($\ell=z$); and
$\epsilon_{\infty,x}=4.95$,
$s_{x 1}=232\,$meV,
$\omega_{x 1}=170\,$meV,
and
$\gamma_{x 1}=3.6\,$meV,
$s_{x 2}=43.5\,$meV,
$\omega_{x 2}=95.1\,$meV,
and
$\gamma_{x 2}=3.4\,$meV
for the $x$ component ($\ell=x$). The latter incorporates hBN phonons as Lorentzians terms.

\acknowledgements
This work has been supported in part by the Spanish MINECO (MAT2017-88492-R and SEV2015- 0522), the ERC (Advanced Grant 789104-eNANO), the European Commission (Graphene Flagship 696656), the Catalan CERCA Program, and the Fundació Privada Cellex. SdV acknowledges financial support through the FPU program from the Spanish MECD.


\begin{thebibliography}{31}
\expandafter\ifx\csname natexlab\endcsname\relax\def\natexlab#1{#1}\fi
\expandafter\ifx\csname bibnamefont\endcsname\relax
  \def\bibnamefont#1{#1}\fi
\expandafter\ifx\csname bibfnamefont\endcsname\relax
  \def\bibfnamefont#1{#1}\fi
\expandafter\ifx\csname citenamefont\endcsname\relax
  \def\citenamefont#1{#1}\fi
\expandafter\ifx\csname url\endcsname\relax
  \def\url#1{\texttt{#1}}\fi
\expandafter\ifx\csname urlprefix\endcsname\relax\def\urlprefix{URL }\fi
\providecommand{\bibinfo}[2]{#2}
\providecommand{\eprint}[2][]{\url{#2}}

\bibitem[{\citenamefont{Schuller et~al.}(2010)\citenamefont{Schuller, Barnard,
  Cai, Jun, White, and Brongersma}}]{SBC10}
\bibinfo{author}{\bibfnamefont{J.~A.} \bibnamefont{Schuller}},
  \bibinfo{author}{\bibfnamefont{E.~S.} \bibnamefont{Barnard}},
  \bibinfo{author}{\bibfnamefont{W.}~\bibnamefont{Cai}},
  \bibinfo{author}{\bibfnamefont{Y.~C.} \bibnamefont{Jun}},
  \bibinfo{author}{\bibfnamefont{J.~S.} \bibnamefont{White}}, \bibnamefont{and}
  \bibinfo{author}{\bibfnamefont{M.~L.} \bibnamefont{Brongersma}},
  \bibinfo{journal}{Nat.\ Mater.} \textbf{\bibinfo{volume}{9}},
  \bibinfo{pages}{193} (\bibinfo{year}{2010}).

\bibitem[{\citenamefont{Halas et~al.}(2011)\citenamefont{Halas, Lal, Chang,
  Link, and Nordlander}}]{HLC11}
\bibinfo{author}{\bibfnamefont{N.~J.} \bibnamefont{Halas}},
  \bibinfo{author}{\bibfnamefont{S.}~\bibnamefont{Lal}},
  \bibinfo{author}{\bibfnamefont{W.}~\bibnamefont{Chang}},
  \bibinfo{author}{\bibfnamefont{S.}~\bibnamefont{Link}}, \bibnamefont{and}
  \bibinfo{author}{\bibfnamefont{P.}~\bibnamefont{Nordlander}},
  \bibinfo{journal}{Chem. Rev.} \textbf{\bibinfo{volume}{111}},
  \bibinfo{pages}{3913} (\bibinfo{year}{2011}).

\bibitem[{\citenamefont{Liz-Marz\'an}(2006)}]{L06}
\bibinfo{author}{\bibfnamefont{L.~M.} \bibnamefont{Liz-Marz\'an}},
  \bibinfo{journal}{Langmuir} \textbf{\bibinfo{volume}{22}},
  \bibinfo{pages}{32} (\bibinfo{year}{2006}).

\bibitem[{\citenamefont{Nagpal et~al.}(2009)\citenamefont{Nagpal, Lindquist,
  Oh, and Norris}}]{NLO09}
\bibinfo{author}{\bibfnamefont{P.}~\bibnamefont{Nagpal}},
  \bibinfo{author}{\bibfnamefont{N.~C.} \bibnamefont{Lindquist}},
  \bibinfo{author}{\bibfnamefont{S.-H.} \bibnamefont{Oh}}, \bibnamefont{and}
  \bibinfo{author}{\bibfnamefont{D.~J.} \bibnamefont{Norris}},
  \bibinfo{journal}{Science} \textbf{\bibinfo{volume}{325}},
  \bibinfo{pages}{594} (\bibinfo{year}{2009}).

\bibitem[{\citenamefont{Zeng et~al.}(2014)\citenamefont{Zeng, Baillargeat, Hod,
  and Yong}}]{ZBH14}
\bibinfo{author}{\bibfnamefont{S.}~\bibnamefont{Zeng}},
  \bibinfo{author}{\bibfnamefont{D.}~\bibnamefont{Baillargeat}},
  \bibinfo{author}{\bibfnamefont{H.-P.} \bibnamefont{Hod}}, \bibnamefont{and}
  \bibinfo{author}{\bibfnamefont{K.-T.} \bibnamefont{Yong}},
  \bibinfo{journal}{Chem.\ Soc.\ Rev.} \textbf{\bibinfo{volume}{43}},
  \bibinfo{pages}{3426} (\bibinfo{year}{2014}).

\bibitem[{\citenamefont{Clavero}(2014)}]{C14}
\bibinfo{author}{\bibfnamefont{C.}~\bibnamefont{Clavero}},
  \bibinfo{journal}{Nat.\ Photon.} \textbf{\bibinfo{volume}{8}},
  \bibinfo{pages}{95} (\bibinfo{year}{2014}).

\bibitem[{\citenamefont{Danckwerts and Novotny}(2007)}]{DN07}
\bibinfo{author}{\bibfnamefont{M.}~\bibnamefont{Danckwerts}} \bibnamefont{and}
  \bibinfo{author}{\bibfnamefont{L.}~\bibnamefont{Novotny}},
  \bibinfo{journal}{Phys.\ Rev.\ Lett.} \textbf{\bibinfo{volume}{98}},
  \bibinfo{pages}{026104} (\bibinfo{year}{2007}).

\bibitem[{\citenamefont{Zia et~al.}(2006)\citenamefont{Zia, Schuller, Chandran,
  and Brongersma}}]{ZSC06}
\bibinfo{author}{\bibfnamefont{R.}~\bibnamefont{Zia}},
  \bibinfo{author}{\bibfnamefont{J.~A.} \bibnamefont{Schuller}},
  \bibinfo{author}{\bibfnamefont{A.}~\bibnamefont{Chandran}}, \bibnamefont{and}
  \bibinfo{author}{\bibfnamefont{M.~L.} \bibnamefont{Brongersma}},
  \bibinfo{journal}{Mater.\ Today} \textbf{\bibinfo{volume}{9}},
  \bibinfo{pages}{20} (\bibinfo{year}{2006}).

\bibitem[{\citenamefont{Ropers et~al.}(2007)\citenamefont{Ropers, Neacsu,
  Elsaesser, Albrecht, Raschke, and Lienau}}]{RNE07}
\bibinfo{author}{\bibfnamefont{C.}~\bibnamefont{Ropers}},
  \bibinfo{author}{\bibfnamefont{C.~C.} \bibnamefont{Neacsu}},
  \bibinfo{author}{\bibfnamefont{T.}~\bibnamefont{Elsaesser}},
  \bibinfo{author}{\bibfnamefont{M.}~\bibnamefont{Albrecht}},
  \bibinfo{author}{\bibfnamefont{M.~B.} \bibnamefont{Raschke}},
  \bibnamefont{and} \bibinfo{author}{\bibfnamefont{C.}~\bibnamefont{Lienau}},
  \bibinfo{journal}{Nano\ Lett.} \textbf{\bibinfo{volume}{7}},
  \bibinfo{pages}{2784} (\bibinfo{year}{2007}).

\bibitem[{\citenamefont{{Alcaraz Iranzo} et~al.}(2018)\citenamefont{{Alcaraz
  Iranzo}, Nanot, Dias, Epstein, Peng, Efetov, Lundeberg, Parret, Osmond, Hong
  et~al.}}]{AND18}
\bibinfo{author}{\bibfnamefont{D.}~\bibnamefont{{Alcaraz Iranzo}}},
  \bibinfo{author}{\bibfnamefont{S.}~\bibnamefont{Nanot}},
  \bibinfo{author}{\bibfnamefont{E.~J.~C.} \bibnamefont{Dias}},
  \bibinfo{author}{\bibfnamefont{I.}~\bibnamefont{Epstein}},
  \bibinfo{author}{\bibfnamefont{C.}~\bibnamefont{Peng}},
  \bibinfo{author}{\bibfnamefont{D.~K.} \bibnamefont{Efetov}},
  \bibinfo{author}{\bibfnamefont{M.~B.} \bibnamefont{Lundeberg}},
  \bibinfo{author}{\bibfnamefont{R.}~\bibnamefont{Parret}},
  \bibinfo{author}{\bibfnamefont{J.}~\bibnamefont{Osmond}},
  \bibinfo{author}{\bibfnamefont{J.-Y.} \bibnamefont{Hong}},
  \bibnamefont{et~al.}, \bibinfo{journal}{Science}
  \textbf{\bibinfo{volume}{360}}, \bibinfo{pages}{291} (\bibinfo{year}{2018}).

\bibitem[{\citenamefont{Fei et~al.}(2012)\citenamefont{Fei, Rodin, Andreev,
  Bao, McLeod, Wagner, Zhang, Zhao, Thiemens, Dominguez et~al.}}]{FRA12}
\bibinfo{author}{\bibfnamefont{Z.}~\bibnamefont{Fei}},
  \bibinfo{author}{\bibfnamefont{A.~S.} \bibnamefont{Rodin}},
  \bibinfo{author}{\bibfnamefont{G.~O.} \bibnamefont{Andreev}},
  \bibinfo{author}{\bibfnamefont{W.}~\bibnamefont{Bao}},
  \bibinfo{author}{\bibfnamefont{A.~S.} \bibnamefont{McLeod}},
  \bibinfo{author}{\bibfnamefont{M.}~\bibnamefont{Wagner}},
  \bibinfo{author}{\bibfnamefont{L.~M.} \bibnamefont{Zhang}},
  \bibinfo{author}{\bibfnamefont{Z.}~\bibnamefont{Zhao}},
  \bibinfo{author}{\bibfnamefont{M.}~\bibnamefont{Thiemens}},
  \bibinfo{author}{\bibfnamefont{G.}~\bibnamefont{Dominguez}},
  \bibnamefont{et~al.}, \bibinfo{journal}{Nature}
  \textbf{\bibinfo{volume}{487}}, \bibinfo{pages}{82} (\bibinfo{year}{2012}).

\bibitem[{\citenamefont{Chen et~al.}(2012)\citenamefont{Chen, Badioli,
  Alonso-Gonz\'alez, Thongrattanasiri, Huth, Osmond, Spasenovi\'c, Centeno,
  Pesquera, Godignon et~al.}}]{paper196}
\bibinfo{author}{\bibfnamefont{J.}~\bibnamefont{Chen}},
  \bibinfo{author}{\bibfnamefont{M.}~\bibnamefont{Badioli}},
  \bibinfo{author}{\bibfnamefont{P.}~\bibnamefont{Alonso-Gonz\'alez}},
  \bibinfo{author}{\bibfnamefont{S.}~\bibnamefont{Thongrattanasiri}},
  \bibinfo{author}{\bibfnamefont{F.}~\bibnamefont{Huth}},
  \bibinfo{author}{\bibfnamefont{J.}~\bibnamefont{Osmond}},
  \bibinfo{author}{\bibfnamefont{M.}~\bibnamefont{Spasenovi\'c}},
  \bibinfo{author}{\bibfnamefont{A.}~\bibnamefont{Centeno}},
  \bibinfo{author}{\bibfnamefont{A.}~\bibnamefont{Pesquera}},
  \bibinfo{author}{\bibfnamefont{P.}~\bibnamefont{Godignon}},
  \bibnamefont{et~al.}, \bibinfo{journal}{Nature}
  \textbf{\bibinfo{volume}{487}}, \bibinfo{pages}{77} (\bibinfo{year}{2012}).

\bibitem[{\citenamefont{Bharadwaj et~al.}(2011)\citenamefont{Bharadwaj,
  Bouhelier, and Novotny}}]{BBN11}
\bibinfo{author}{\bibfnamefont{P.}~\bibnamefont{Bharadwaj}},
  \bibinfo{author}{\bibfnamefont{A.}~\bibnamefont{Bouhelier}},
  \bibnamefont{and} \bibinfo{author}{\bibfnamefont{L.}~\bibnamefont{Novotny}},
  \bibinfo{journal}{Phys.\ Rev.\ Lett.} \textbf{\bibinfo{volume}{106}},
  \bibinfo{pages}{226802} (\bibinfo{year}{2011}).

\bibitem[{\citenamefont{Nikitin et~al.}(2011)\citenamefont{Nikitin, Guinea,
  Garc\'{\i}a-Vidal, and Mart\'{\i}n-Moreno}}]{NGG11_2}
\bibinfo{author}{\bibfnamefont{A.~Y.} \bibnamefont{Nikitin}},
  \bibinfo{author}{\bibfnamefont{F.}~\bibnamefont{Guinea}},
  \bibinfo{author}{\bibfnamefont{F.~J.} \bibnamefont{Garc\'{\i}a-Vidal}},
  \bibnamefont{and}
  \bibinfo{author}{\bibfnamefont{L.}~\bibnamefont{Mart\'{\i}n-Moreno}},
  \bibinfo{journal}{Phys.\ Rev.\ B} \textbf{\bibinfo{volume}{84}},
  \bibinfo{pages}{195446} (\bibinfo{year}{2011}).

\bibitem[{\citenamefont{Ritchie}(1957)}]{R1957}
\bibinfo{author}{\bibfnamefont{R.~H.} \bibnamefont{Ritchie}},
  \bibinfo{journal}{Phys.\ Rev.} \textbf{\bibinfo{volume}{106}},
  \bibinfo{pages}{874} (\bibinfo{year}{1957}).

\bibitem[{\citenamefont{Powell and Swan}(1959)}]{PS1959}
\bibinfo{author}{\bibfnamefont{C.~J.} \bibnamefont{Powell}} \bibnamefont{and}
  \bibinfo{author}{\bibfnamefont{J.~B.} \bibnamefont{Swan}},
  \bibinfo{journal}{Phys.\ Rev.} \textbf{\bibinfo{volume}{115}},
  \bibinfo{pages}{869} (\bibinfo{year}{1959}).

\bibitem[{\citenamefont{{Garc\'{\i}a de Abajo}}(2010)}]{paper149}
\bibinfo{author}{\bibfnamefont{F.~J.} \bibnamefont{{Garc\'{\i}a de Abajo}}},
  \bibinfo{journal}{Rev.\ Mod.\ Phys.} \textbf{\bibinfo{volume}{82}},
  \bibinfo{pages}{209} (\bibinfo{year}{2010}).

\bibitem[{\citenamefont{Persson and Baratoff}(1992)}]{PB92}
\bibinfo{author}{\bibfnamefont{B.~N.~J.} \bibnamefont{Persson}}
  \bibnamefont{and} \bibinfo{author}{\bibfnamefont{A.}~\bibnamefont{Baratoff}},
  \bibinfo{journal}{Phys.\ Rev.\ Lett.} \textbf{\bibinfo{volume}{68}},
  \bibinfo{pages}{3224} (\bibinfo{year}{1992}).

\bibitem[{\citenamefont{Parzefall et~al.}(2015)\citenamefont{Parzefall,
  Bharadwaj, Jain, Taniguchi, Watanabe, and Novotny}}]{PPJ15}
\bibinfo{author}{\bibfnamefont{M.}~\bibnamefont{Parzefall}},
  \bibinfo{author}{\bibfnamefont{P.}~\bibnamefont{Bharadwaj}},
  \bibinfo{author}{\bibfnamefont{A.}~\bibnamefont{Jain}},
  \bibinfo{author}{\bibfnamefont{T.}~\bibnamefont{Taniguchi}},
  \bibinfo{author}{\bibfnamefont{K.}~\bibnamefont{Watanabe}}, \bibnamefont{and}
  \bibinfo{author}{\bibfnamefont{L.}~\bibnamefont{Novotny}},
  \bibinfo{journal}{Nat.\ Nanotech.} \textbf{\bibinfo{volume}{0}},
  \bibinfo{pages}{DOI: 10.1038/NNANO.2015.203} (\bibinfo{year}{2015}).

\bibitem[{\citenamefont{Walters et~al.}(2010)\citenamefont{Walters, {van Loon},
  Brunets, Schmitz, and Polman}}]{WVB10}
\bibinfo{author}{\bibfnamefont{R.~J.} \bibnamefont{Walters}},
  \bibinfo{author}{\bibfnamefont{R.~V.~A.} \bibnamefont{{van Loon}}},
  \bibinfo{author}{\bibfnamefont{I.}~\bibnamefont{Brunets}},
  \bibinfo{author}{\bibfnamefont{J.}~\bibnamefont{Schmitz}}, \bibnamefont{and}
  \bibinfo{author}{\bibfnamefont{A.}~\bibnamefont{Polman}},
  \bibinfo{journal}{Nat.\ Mater.} \textbf{\bibinfo{volume}{9}},
  \bibinfo{pages}{21} (\bibinfo{year}{2010}).

\bibitem[{\citenamefont{Zhang et~al.}(2013)\citenamefont{Zhang, Allen, and
  DeCorby}}]{ZAD13}
\bibinfo{author}{\bibfnamefont{M.~C.} \bibnamefont{Zhang}},
  \bibinfo{author}{\bibfnamefont{T.~W.} \bibnamefont{Allen}}, \bibnamefont{and}
  \bibinfo{author}{\bibfnamefont{R.~G.} \bibnamefont{DeCorby}},
  \bibinfo{journal}{Appl.\ Phys.\ Lett.} \textbf{\bibinfo{volume}{103}},
  \bibinfo{pages}{071109} (\bibinfo{year}{2013}).

\bibitem[{\citenamefont{Svintson et~al.}(2016)\citenamefont{Svintson,
  Devizorova, Otsuji, and Ryzhii}}]{SDO16}
\bibinfo{author}{\bibfnamefont{D.}~\bibnamefont{Svintson}},
  \bibinfo{author}{\bibfnamefont{Z.}~\bibnamefont{Devizorova}},
  \bibinfo{author}{\bibfnamefont{T.}~\bibnamefont{Otsuji}}, \bibnamefont{and}
  \bibinfo{author}{\bibfnamefont{V.}~\bibnamefont{Ryzhii}},
  \bibinfo{journal}{Phys.\ Rev.\ B} \textbf{\bibinfo{volume}{94}},
  \bibinfo{pages}{115301} (\bibinfo{year}{2016}).

\bibitem[{\citenamefont{Svintsov et~al.}(2016)\citenamefont{Svintsov,
  Devizorova, Otsuji, and Ryzhii}}]{SDR16}
\bibinfo{author}{\bibfnamefont{D.}~\bibnamefont{Svintsov}},
  \bibinfo{author}{\bibfnamefont{Z.}~\bibnamefont{Devizorova}},
  \bibinfo{author}{\bibfnamefont{T.}~\bibnamefont{Otsuji}}, \bibnamefont{and}
  \bibinfo{author}{\bibfnamefont{V.}~\bibnamefont{Ryzhii}},
  \bibinfo{journal}{Phys.\ Rev.\ B} \textbf{\bibinfo{volume}{94}},
  \bibinfo{pages}{115301} (\bibinfo{year}{2016}).

\bibitem[{\citenamefont{Woessner et~al.}(2017)\citenamefont{Woessner, Misra,
  Cao, Torre, Mishchenko, Lundeberg, Watanabe, Taniguchi, Polini, Novoselov
  et~al.}}]{WMC17}
\bibinfo{author}{\bibfnamefont{A.}~\bibnamefont{Woessner}},
  \bibinfo{author}{\bibfnamefont{A.}~\bibnamefont{Misra}},
  \bibinfo{author}{\bibfnamefont{Y.}~\bibnamefont{Cao}},
  \bibinfo{author}{\bibfnamefont{I.}~\bibnamefont{Torre}},
  \bibinfo{author}{\bibfnamefont{A.}~\bibnamefont{Mishchenko}},
  \bibinfo{author}{\bibfnamefont{M.~B.} \bibnamefont{Lundeberg}},
  \bibinfo{author}{\bibfnamefont{K.}~\bibnamefont{Watanabe}},
  \bibinfo{author}{\bibfnamefont{T.}~\bibnamefont{Taniguchi}},
  \bibinfo{author}{\bibfnamefont{M.}~\bibnamefont{Polini}},
  \bibinfo{author}{\bibfnamefont{K.~S.} \bibnamefont{Novoselov}},
  \bibnamefont{et~al.}, \bibinfo{journal}{ACS\ Photon.}
  \textbf{\bibinfo{volume}{4}}, \bibinfo{pages}{1461} (\bibinfo{year}{2017}).

\bibitem[{\citenamefont{de~Vega and {Garc\'{\i}a de Abajo}}(2017)}]{paper295}
\bibinfo{author}{\bibfnamefont{S.}~\bibnamefont{de~Vega}} \bibnamefont{and}
  \bibinfo{author}{\bibfnamefont{F.~J.} \bibnamefont{{Garc\'{\i}a de Abajo}}},
  \bibinfo{journal}{ACS\ Photon.} \textbf{\bibinfo{volume}{4}},
  \bibinfo{pages}{2367} (\bibinfo{year}{2017}).

\bibitem[{\citenamefont{{Castro Neto} et~al.}(2009)\citenamefont{{Castro Neto},
  Guinea, Peres, Novoselov, and Geim}}]{CGP09}
\bibinfo{author}{\bibfnamefont{A.~H.} \bibnamefont{{Castro Neto}}},
  \bibinfo{author}{\bibfnamefont{F.}~\bibnamefont{Guinea}},
  \bibinfo{author}{\bibfnamefont{N.~M.~R.} \bibnamefont{Peres}},
  \bibinfo{author}{\bibfnamefont{K.~S.} \bibnamefont{Novoselov}},
  \bibnamefont{and} \bibinfo{author}{\bibfnamefont{A.~K.} \bibnamefont{Geim}},
  \bibinfo{journal}{Rev.\ Mod.\ Phys.} \textbf{\bibinfo{volume}{81}},
  \bibinfo{pages}{109} (\bibinfo{year}{2009}).

\bibitem[{\citenamefont{Wunsch et~al.}(2006)\citenamefont{Wunsch, Stauber,
  Sols, and Guinea}}]{WSS06}
\bibinfo{author}{\bibfnamefont{B.}~\bibnamefont{Wunsch}},
  \bibinfo{author}{\bibfnamefont{T.}~\bibnamefont{Stauber}},
  \bibinfo{author}{\bibfnamefont{F.}~\bibnamefont{Sols}}, \bibnamefont{and}
  \bibinfo{author}{\bibfnamefont{F.}~\bibnamefont{Guinea}},
  \bibinfo{journal}{New\ J.\ Phys.} \textbf{\bibinfo{volume}{8}},
  \bibinfo{pages}{318} (\bibinfo{year}{2006}).

\bibitem[{\citenamefont{Hwang and {Das Sarma}}(2007)}]{HD07}
\bibinfo{author}{\bibfnamefont{E.~H.} \bibnamefont{Hwang}} \bibnamefont{and}
  \bibinfo{author}{\bibfnamefont{S.}~\bibnamefont{{Das Sarma}}},
  \bibinfo{journal}{Phys.\ Rev.\ B} \textbf{\bibinfo{volume}{75}},
  \bibinfo{pages}{205418} (\bibinfo{year}{2007}).

\bibitem[{\citenamefont{Geick et~al.}(1966)\citenamefont{Geick, Perry, and
  Rupprecht}}]{GPR1966}
\bibinfo{author}{\bibfnamefont{R.}~\bibnamefont{Geick}},
  \bibinfo{author}{\bibfnamefont{C.~H.} \bibnamefont{Perry}}, \bibnamefont{and}
  \bibinfo{author}{\bibfnamefont{G.}~\bibnamefont{Rupprecht}},
  \bibinfo{journal}{Phys.\ Rev.} \textbf{\bibinfo{volume}{146}},
  \bibinfo{pages}{543} (\bibinfo{year}{1966}).

\bibitem[{\citenamefont{Palik}(1985)}]{P1985}
\bibinfo{author}{\bibfnamefont{E.~D.} \bibnamefont{Palik}},
  \emph{\bibinfo{title}{Handbook of Optical Constants of Solids}}
  (\bibinfo{publisher}{Academic Press}, \bibinfo{address}{San Diego},
  \bibinfo{year}{1985}).

\bibitem[{\citenamefont{Filatova and Konashuk}(2015)}]{FK15}
\bibinfo{author}{\bibfnamefont{E.~O.} \bibnamefont{Filatova}} \bibnamefont{and}
  \bibinfo{author}{\bibfnamefont{A.~S.} \bibnamefont{Konashuk}},
  \bibinfo{journal}{J.\ Phys.\ Chem.\ C} \textbf{\bibinfo{volume}{119}},
  \bibinfo{pages}{20755} (\bibinfo{year}{2015}).

\end{thebibliography}

\end{document}